\documentclass[aps]{revtex4}

\begin{document}
\title{Infrared behavior and fixed-point structure in the compactified Ginzburg--Landau model}
\author{C.A. Linhares$^{(a)}$, A.P.C. Malbouisson$^{(b)}$ and M.L. Souza$^{(b)}$}
\address{$^{(a)}$Instituto de F\'{\i}sica, Universidade do Estado do Rio de Janeiro,\\
Rua S\~{a}o Francisco Xavier, 524, 20559-900 Rio de Janeiro, RJ, Brazil\\
$^{(b)}$Centro Brasileiro de Pesquisas F\'{\i}sicas/MCT, Rua Dr. Xavier\\
Sigaud, 150, 22290-180 Rio de Janeiro, RJ, Brazil}

\begin{abstract}
We consider the Euclidean $N$-component Ginzburg--Landau model in $D$
dimensions, of which $d$ ($d\leq D$) of them are compactified. As usual,
temperature is introduced through the mass term in the Hamiltonian. 
This model can be interpreted as describing a system in a region of the $D$-dimensional space, limited by $d$ pairs of parallel planes, orthogonal to the
coordinates axis $x_1,\,x_2,\,\ldots ,\,x_d$. The planes in each
pair are separated by distances $L_1,\;L_2,\;\ldots ,\,L_d$. For $D=3$, from a physical point of view, the system can
be supposed to describe, in the cases of $d=1$, $d=2$, and $d=3$,
respectively, a superconducting material in the form of a film, of an
infinitely long wire having a retangular cross-section and of a brick-shaped
grain. We investigate in the large-$N$ limit the fixed-point structure of
the model, in the absence or presence of an external magnetic field. An
infrared-stable fixed point is found, whether of not an external magnetic
field is applied, but for different ranges of values of the space dimension $
D$.
\end{abstract}

\maketitle
\address{$^{(a)}$Instituto de F\'{\i}sica, Universidade do Estado do Rio de Janeiro,\\
Rua S\~{a}o Francisco Xavier, 524, 20559-900 Rio de Janeiro, RJ, Brazil\\
$^{(b)}$Centro Brasileiro de Pesquisas F\'{\i}sicas/MCT, Rua Dr. Xavier\\
Sigaud, 150, 22290-180 Rio de Janeiro, RJ, Brazil}
\address{$^{(a)}$Instituto de F\'{\i}sica, Universidade do Estado do Rio de Janeiro,\\
Rua S\~{a}o Francisco Xavier, 524, 20559-900 Rio de Janeiro, RJ, Brazil\\
$^{(b)}$Centro Brasileiro de Pesquisas F\'{\i}sicas/MCT, Rua Dr. Xavier\\
Sigaud, 150, 22290-180 Rio de Janeiro, RJ, Brazil}

\address{$^{(a)}$Instituto de F\'{\i}sica, Universidade do Estado do Rio de Janeiro,\\
Rua S\~{a}o Francisco Xavier, 524, 20559-900 Rio de Janeiro, RJ, Brazil\\
$^{(b)}$Centro Brasileiro de Pesquisas F\'{\i}sicas/MCT, Rua Dr. Xavier\\
Sigaud, 150, 22290-180 Rio de Janeiro, RJ, Brazil}

\section{Introduction}

A large amount of work has already been done on the Ginzburg--Landau (GL)
model, both in its single component and in the $N$-component versions, using
the renormalization group approach~\cite
{affleck,lawrie,lawrie1,brezin,radz,flavio,malbo}. In particular, an
analysis of the renormalization group in finite-size geometries can be found
in~\cite{zinn,cardy} and a general study of phase transitions in confined
systems is in~\cite{livro}. These studies have been performed to take into
account boundary effects on thermodynamical quantities for these systems.
The existence of phase transitions are in this case associated to some
spatial parameters related to the breaking of translational invariance, for
instance, the distance $L$ between planes confining the system. Also, in
other contexts, the influence of boundaries in the behavior of systems
undergoing transitions have been investigated~\cite{fosco,fadolfo}.

We shall analyze in the present paper the effects of boundaries on the
transition by considering that such confined systems are modeled by
compactifying spatial dimensions~\cite{livro}. Compactification will be
engendered as a generalization of the Matsubara (imaginary-time)
prescription to account for constraints on the spatial coordinates. In the
original Matsubara formalism, time is rotated to the imaginary axis, $%
t\rightarrow i\tau $, where $\tau $ (the Euclidean time) is limited to the
interval $0\leq \tau \leq \beta $, with $\beta =1/T$ standing for the
inverse temperature. The fields then fulfill periodic (bosons) or
antiperiodic (fermions) boundary conditions and are compactified on the $%
\tau $-axis in an $S^1$ topology, the circumference of length $\beta $. Such
a formalism leads to the description of a system in thermal equilibrium at
the temperature $\beta ^{-1}$. Since in a Euclidean field theory space and
time are on the same footing, one can envisage a generalization of the
Matsubara approach to any set of spatial coordinates as well~\cite
{polchinski,atickwitten,tashk,ademir}.

The topological conceptual framework for studying simultaneously finite
temperature and spatial constraints has been developed by considering a
simply or nonsimply connected $D$-dimensional manifold with a topology of
the type $\Gamma _D^{d+1}={\rm{\bf R}}^{D-d-1}\times {\rm{\bf S} }^{1_0}\times {\rm{\bf S} 
}^{1_1}\times \cdots \times {\rm{\bf S} }^{1_d}$, with ${\rm{\bf S} }^{1_0}$
corresponding to the compactification of the imaginary time and $\,{\rm{\bf S} }
^{1_1},\dots ,{\rm{\bf S} }^{1_d}$ referring to the compactification of $d$
spatial dimensions~\cite{ademir,khanna1}. The topological structure of
spacetime does not modify the local field equations. However, topology
implies modifications of the boundary conditions on fields and Green
functions~\cite{ford1}. Physical manifestations of this type of topology
include, for instance, the vacuum-energy fluctuations giving rise to the
Casimir effect~\cite{livro,hebe,jura1,mostep3}; in the study of phase
transitions, the dependence of the critical temperature on the
compactification parameters is found in several situations of
condensed-matter physics~\cite{livro,amms,linhares,linhares1,jmp,jmp1}.
Also, this kind of formalism has been employed in the investigation of the
confining phase transition in effective theories for Quantum Chromodynamics~%
\cite{prd11,epl10,gnn,gnn2,npb09}. In the $\Gamma _D^{d+1}$topology, the
Feynman rules are modified by introducing a generalized Matsubara
prescription, performing the following multiple replacements
[compactification of a $(d+1)$-dimensional subspace]: 
\begin{equation}
\int \frac{dk_0}{2\pi }\rightarrow \frac 1\beta \sum_{n_1=-\infty }^{+\infty
}\,,\;\;\;\;\int \frac{dk_i}{2\pi }\rightarrow \frac 1{L_i}\sum_{n_i=-\infty
}^{+\infty }\;;\;\;\;\;k_1\rightarrow \frac{2(n_1+c)\pi }\beta
\;\;\;k_i\rightarrow \frac{2(n_i+c)\pi }{L_i}\;,  \label{Matsubara1}
\end{equation}
where for each $i=1,2,\ldots ,d$, $L_i\,$is the size of the compactified
spatial dimension $i$ and $c=0$ or $c=1/2$ for, respectively, bosons and
fermions.

The compactification formalism described above has been applied to
field-theoretical models in $D$ dimensions, with a $d$-dimensional ($d\leq D$%
) set of compactified spatial coordinates~\cite{jmp,jmp1,jmario}. This
formalism has also been developed from a path-integral approach in~\cite
{khanna1}. This allows to generalize to any subspace previous results in the
effective potential framework for finite temperature and spatial boundaries.
This mechanism generalizes and unifies results from recent work on the
behavior of field theories in the presence of spatial constraints~\cite
{ademir,jmario,fadolfo}, and previous results in the literature for
finite-temperature field theory as, for instance, in~\cite{gino}.

When studying the compactification of spatial coordinates, however, it is
argued in \cite{livro} from topological considerations, that we may have a
quite different interpretation of the generalized Matsubara prescription: it
provides a general and practical way to account for systems confined in
limited regions of space at finite temperatures. Distinctly, we shall be
concerned here with stationary field theories and employ the generalized
Matsubara prescription to study bounded systems by implementing the
compactification of spatial coordinates; no imaginary-time compactification
will be done, temperature will be introduced through the mass parameter in
the Hamiltonian. We will consider a topology of the type $\Gamma _D^d={\rm{\bf  R} 
}^{D-d}\times {\rm{\bf S} }^{1_1}\times {\rm{\bf S} }^{1_2}\times \cdots \times {\rm{\bf S} 
}^{1_d}$, where $\,{\rm{\bf S} }^{1_1},\dots ,{\rm{\bf S} }^{1_d}$ refer to the
compactification of $d$ {\em spatial} dimensions.

We consider in the present article the Euclidean vector $N$-component $
(\lambda \varphi ^4)_D$ theory at leading order in $1/N$, the system being
submitted to the constraint of being limited by $d$ pairs of parallel
planes. Each pair is orthogonal to the coordinate axes $x_1,\ldots ,x_d$,
respectively, and in each one of them the planes are at distances $%
L_1,\ldots ,L_d$ apart from one another. From a physical point of view, we
take in particular $D=3$ and introduce temperature by means of the mass term
in the Hamiltonian in the usual Ginzburg--Landau fashion. These models can
then describe a superconducting material in the shapes of a film ($d=1$), of
a wire ($d=2$) and of a grain ($d=3$). With geometries such as these, some
of us have been able to obtain general formulas for the dependence of the
transition temperature and other quantities on the parameters delimiting the
spatial region within which the system is confined (see for instance~\cite
{jmp,jmp1} and other references therein).

We also consider the critical behavior of the system under the influence of
an external magnetic field. Physically, for $D=3$, this corresponds to
superconducting films, wires and grains in a magnetic field. In~\cite{radz},
a large-$N$ theory of a second-order transition for arbitrary dimension $D$
is presented and the fixed-point effective free energy describing the
transition is found. The theory is based on the Ginzburg--Landau model with
the coupling of scalar and gauge fields. While ignoring gauge-field
fluctuations, the model includes an external magnetic field. The authors in~%
\cite{radz} also claim that it is possible that in the physical situation of 
$N=1$, a mechanism of reduction of the lower critical dimension could allow
a continuous transition in $D=3$. In \cite{malbo}, the possibility of the
existence of a phase transition for a superconductor film in the presence of
an external magnetic field has been investigated. This has been done in the
renormalization-group framework by looking for the existence of
infrared-stable fixed points for the $\beta $ function.

In this article, we study, for arbitrary space dimension $D$ and for any
number $d\leq D$ of compactified dimensions (specially wires and grains),
the fixed-point structure of the model, thus generalizing the previously
quoted studies for films. In both situations, with or without
external magnetic field, we shall neglect the minimal coupling with the
vector potential corresponding to the intrinsic gauge fluctuations. Also, as
usual in the GL model, no imaginary-time compactification will be done,
temperature will be introduced through the mass parameter in the Hamiltonian.
Our main concern will be to analyze the model from
a field-theoretical point of view. In this sense, the present work may be seen as a further development of previous papers by some of us, as for instance~\cite{malbo,ademir,khanna1}.

The paper is organized in the following way. In Section II below, we
establish in all compactified cases the running coupling constant (and hence
the fixed point) for the model in which the external field is omitted, while
the analogous study when it is considered is the subject of Section III. In
Section IV, we present our conclusions.

\section{The compactified model in the absence of an external field}

We first consider the $N$-component vector model described by the
Ginzburg--Landau Hamiltonian density 
\begin{equation}
{\cal H}=\partial _\mu \varphi _a\partial ^\mu \varphi _a+m^2\varphi
_a\varphi _a+u\,(\varphi _a\varphi _a)^2\,  \label{hamiltoniana}
\end{equation}
in Euclidean $D$-dimensional space, where $u$ is the coupling constant and $
m^2$ is a mass parameter such that $m^2=\alpha \left( T-T_0\right) $ and $
T_0 $ the bulk transition temperature. Summation over repeated indices $\mu $
and $a$ is assumed. In the following, we will consider the model described
by the Hamiltonian (\ref{hamiltoniana}) and take the large-$N$ limit, such
that $u\rightarrow 0$, $N\rightarrow \infty $ with $Nu=\lambda $ fixed.

Let us consider the system in $D$ dimensions confined to a region of space
delimited by $d$ $(d\leq D)$ pairs of parallel planes. Each plane of a pair $j$ is at
a distance $L_j$ from the other member of the pair, $j=1,2,\ldots ,d$, and
is orthogonal to all other planes belonging to distinct pairs $i$, $i\neq j$%
. This may be pictured as a parallelepiped-shaped box embedded in the $D$%
-dimensional space, whose parallel faces are separated by distances $L_1$, $%
L_2$,$\ldots $, $L_d$. We use Cartesian coordinates ${\bf r}=(x_1,\ldots
,x_d,{\bf z})$, where ${\bf z}$ is a $(D-d)$-dimensional vector, with
corresponding momenta ${\bf k}=(k_1,\ldots ,k_d,{\bf q})$, ${\bf q}$ being a 
$(D-d)$-dimensional vector in momentum space. Under these conditions, the
generating functional of correlation functions is written in the form 
\begin{equation}
{\cal Z}=\int {\cal D}\varphi ^{*}{\cal D}\varphi \,\exp \left(
-\int_0^{L_1}dx_1\cdots \int_0^{L_d}dx_d\int d^{D-d}z\;{\cal H}(|\varphi
|,|\nabla \varphi |\right) ,  \label{Z}
\end{equation}
with the field $\varphi (x_1,\ldots ,x_d,{\bf z})$ satisfying the condition
of confinement inside the box, $\varphi (x_i\leq 0,{\bf z})\;=\;\varphi
(x\geq L,{\bf z})\;=\;$ const. Then the field should have a mixed
series-integral Fourier expansion of the form 
\begin{equation}
\varphi (x_1,\ldots ,x_d,{\bf z})=\sum_{i=1}^d\sum_{n_i=-\infty }^\infty
c_{n_i}\int d^{D-d}{\bf q}\;b({\bf q})e^{-i\omega _{n_i}x\;-i{\bf q}\cdot 
{\bf z}}\tilde{\varphi}(\omega _{n_i},{\bf q}),  \label{Fourier}
\end{equation}
where, for $i=1,\ldots ,d$, $\omega _{n_i}=2\pi n_i/L_i$ and the
coefficients $c_{n_i}$ and $b({\bf q})$ correspond respectively to the
Fourier series representation over the $x_i$ and to the Fourier integral
representation over the $(D-d)$-dimensional ${\bf z}$-space. As explained in
the comments leading to Eq.~(\ref{Matsubara1}), the above conditions of
confinement of the $x_i$-dependence of the field to a segment of length $L_i$
allow us to proceed with respect to the $x_i$-coordinates, for all $i$, in a
manner analogous as it is done in the imaginary-time Matsubara formalism in
field theory. Accordingly, the multiple Matsubara replacements modify the
Feynman rules following the prescription 
\begin{equation}
\int \frac{dk_i}{2\pi }\rightarrow \frac 1{L_i}\sum_{n_i=-\infty }^{+\infty
},\qquad k_i\rightarrow \frac{2\pi n_i}{L_i}\equiv \omega _{n_i},\qquad
i=1,\ldots ,d.  \label{Matsubara}
\end{equation}
Compactification can be implemented in different ways as, for instance,
through specific conditions on the fields at spatial boundaries. We here
choose periodic boundary conditions.

\subsection{The boundary-dependent coupling constant in the large-$N$ limit}

The coupling constant will be defined in terms of the four-point function
for small external momenta which, at leading order in $1/N$, is given by the
sum of all chains of one-loop diagrams. It is given in momentum space,
before compactification, and at the critical point by \cite{flavio} 
\begin{equation}
\Gamma _D^{(4)}(p,m=0)=\frac u{1+Nu\Pi (p,m=0)}\,,  \label{R1}
\end{equation}
where $\Pi (p,m=0)$ is the single one-loop integral at the critical point.
It is written as (let us keep in mind that $p$ is the $D$-dimensional
external momentum vector) 
\begin{eqnarray}
\Pi \left( p,m=0\right) &=&\int \frac{d^Dk}{(2\pi )^D}\frac 1{\left[
k^2\left( p-k\right) ^2\right] }  \nonumber \\
&=&\int_0^1dx\int \frac{d^Dk}{(2\pi )^D}\frac 1{\left[ k^2+p^2x(1-x)\right]
^2},  \label{Pi}
\end{eqnarray}
where a Feynman parameter $x$ was introduced.

Performing the Matsubara replacements (\ref{Matsubara}) for $d$ dimensions,
Eq. (\ref{Pi}) becomes 
\begin{eqnarray}
\Pi (p,D,\{L_i\},m=0) &=&\frac 1{L_1\cdots L_d}\sum_{i=1}^d\sum_{n_i=-\infty
}^\infty \int_0^1dx\int \frac{d^{D-d}q}{(2\pi )^{D-d}}  \nonumber
\label{sigma1} \\
&&\times \frac 1{\left[ {\bf q}^2+\omega _{n_1}^2+\cdots +\omega
_{n_d}^2+p^2x(1-x)\right] ^2}  \nonumber \\
&&  \label{Pi2}
\end{eqnarray}
and we define the effective $\{L_i\}$-dependent coupling constant in the
large-$N$ limit as 
\begin{equation}
\lambda (p,D,\{L_i\})\equiv \lim_{u\rightarrow 0\,;\;\,N\rightarrow \infty
}N\Gamma _D^{(4)}(p,\{L_i\},m=0)=\frac \lambda {1+\lambda \Pi
(p,D,\{L_i\},m=0)},  \label{lambda}
\end{equation}
with $Nu=\lambda $ fixed.

The sum over the $n_i$ and the integral over ${\bf q}$ above can be treated
using the formalism developed in \cite{ademir}. It concerns the study of
expressions of the form 
\begin{equation}
I(s)=\sum_{i=1}^d\sum_{n_i=-\infty }^{+\infty }\int \frac{d^{D-d}q}{({\bf q}%
^2+a_1n_1^2+\cdots +a_dn_d^2+c^2)^s}.  \label{integral1}
\end{equation}
(In our case, for the computation of $\Pi $, we have $s=2$, $a_i=1/L_i^2$, $%
\omega _i^2=(2\pi )^2a_in_i^2$ and $c^2=p^2x(1-x)/(2\pi )^2$; also, a
redefinition of the integration variables, ${\bf q}\rightarrow {\bf q}/2\pi $%
, has been performed.) Such integral over the $D-d$ noncompactified momentum
variables is performed using the well-known dimensional regularization
formula \cite{zinn} 
\begin{equation}
\int \frac{d^\ell q}{\left( {\bf q}^2+M\right) ^s}=\frac{\Gamma \left( s-%
\frac \ell 2\right) }{\Gamma (s)}\frac{\pi ^{\ell /2}}{M^{s-\ell /2}},
\end{equation}
which, for $\ell =D-d$, leads to 
\begin{equation}
I(s)=f(D,d,s)Z_d^{c^2}\left( s-\frac{D-d}2;a_1,\ldots ,a_d\right) ,
\label{integral2}
\end{equation}
where 
\begin{equation}
f(D,d,s)=\pi ^{(D-d)/2}\frac{\Gamma \left( s-\frac{D-d}2\right) }{\Gamma (s)}
\end{equation}
and $Z_d^{c^2}(\nu ;a_1,\ldots ,a_d)$ are Epstein--Hurwitz zeta functions,
for $\nu =s-(D-d)/2$, which are defined by 
\begin{equation}
Z_d^{c^2}(\nu ;a_1,...,a_d)=\sum_{n_1,...,n_d=-\infty }^\infty
(a_1n_1^2+\cdots +a_dn_d^2+c^2)^{-\nu }.  \label{zeta}
\end{equation}
It is valid for Re$(\nu )>d/2$ (in our case, this implies Re$(s)>D/2$). The
Epstein--Hurwitz zeta function can be extended to the whole complex $s$%
-plane and we obtain, after some manipulations \cite{ademir,elizalde}, 
\begin{eqnarray}
Z_d^{c^2}(\nu ;a_1,...,a_d) &=&\frac{2^{\nu -\frac d2+1}\pi ^{2\nu -\frac d2}%
}{\sqrt{a_1\cdots a_d}\,\Gamma (\nu )}\left[ 2^{\nu -\frac d2-1}c^{d-2\nu
}\Gamma \left( \nu -\frac d2\right) \right.  \nonumber  \label{zeta4} \\
&&\left. +2\sum_{i=1}^d\sum_{n_i=1}^\infty \left( \frac c{L_in_i}\right) ^{%
\frac d2-\nu }K_{\nu -\frac d2}\left( cL_in_i\right) +\cdots \right. 
\nonumber \\
&&\left. +2^d\sum_{n_1,...,n_d=1}^\infty \left( \frac c{\sqrt{%
L_1^2n_1^2+\cdots +L_d^2n_d^2}}\right) ^{\frac d2-\nu }K_{\nu -\frac d2%
}\left( c\sqrt{L_1^2n_1^2+\cdots +L_d^2n_d^2}\right) \right] .  \nonumber \\
&&  \label{zeta4}
\end{eqnarray}
Putting $\nu =s-(D-d)/2$ in Eq. (\ref{zeta4}), we get 
\begin{eqnarray}
I(s) &=&\frac{h(D,s)}{\sqrt{a_1\cdots a_d}}\left[ 2^{s-D/2-2}c^{D-2s}\Gamma
\left( s-\frac D2\right) \right.  \nonumber  \label{potefet3} \\
&&+\sum_{i=1}^d\sum_{n_i=1}^\infty \left( \frac c{L_in_i}\right)
^{D/2-s}K_{D/2-s}(cL_in_i)  \nonumber \\
&&+2\sum_{i<j=1}^d\sum_{n_i,n_j=1}^\infty \left( \frac c{\sqrt{%
L_i^2n_i^2+L_j^2n_j^2}}\right) ^{D/2-s}K_{D/2-s}\left( c\sqrt{%
L_i^2n_i^2+L_j^2n_j^2}\right) +\cdots  \nonumber \\
&&\left. +2^{d-1}\sum_{n_1,\ldots ,n_d=1}^\infty \left( \frac c{\sqrt{%
L_1^2n_1^2+\cdots +L_d^2n_d^2}}\right) ^{D/2-s}K_{D/2-s}\left( c\sqrt{%
L_1^2n_1^2+\cdots +L_d^2n_d^2}\right) \right] ,  \nonumber \\
&&  \label{integral3}
\end{eqnarray}
where 
\begin{equation}
h(D,s)=\frac{2^{s-D/2+2}\pi ^{2s-D/2}}{\Gamma (s)}  \label{h}
\end{equation}
and the $K_\nu $ are the modified Bessel functions. Applying formula (\ref
{integral3}) to Eq. (\ref{Pi}) the result is 
\begin{eqnarray}
\Pi (p,D,\{L_i\},m=0) &=&\frac{\sqrt{a_1\cdots a_d}}{\left( 2\pi \right) ^4}%
\int_0^1dx\,I(2)  \nonumber \\
&=&\frac{h(D,2)}{\left( 2\pi \right) ^4}\int_0^1dx\,\left[ 2^{-D/2}\left( 
\frac 1{\left( 2\pi \right) ^2}p^2x(1-x)\right) ^{D/2-2}\Gamma \left( 2-%
\frac D2\right) \right.  \nonumber \\
&&+\sum_{i=1}^d\sum_{n_i=1}^\infty \left( \frac{\sqrt{p^2x(1-x)}}{2\pi L_in_i%
}\right) ^{D/2-2}K_{D/2-2}\left( \frac 1{2\pi }\sqrt{p^2x(1-x)}L_in_i\right)
\nonumber \\
&&+2\sum_{i<j=1}^d\sum_{n_i,n_j=1}^\infty \left( \frac{\sqrt{p^2x(1-x)}}{%
2\pi \sqrt{L_i^2n_i^2+L_j^2n_j^2}}\right) ^{D/2-2}K_{D/2-2}\left( \frac 1{%
2\pi }\sqrt{p^2x(1-x)}\sqrt{L_i^2n_i^2+L_j^2n_j^2}\right) +\cdots  \nonumber
\\
&&+2^{d-1}\sum_{n_1,\ldots ,n_d=1}^\infty \left( \frac{\sqrt{p^2x(1-x)}}{%
2\pi \sqrt{L_1^2n_1^2+\cdots +L_d^2n_d^2}}\right) ^{D/2-2}  \nonumber \\
&&\left. \qquad \qquad \qquad \qquad \times K_{D/2-2}\left( \frac 1{2\pi }%
\sqrt{p^2x(1-x)}\sqrt{L_1^2n_1^2+\cdots +L_d^2n_d^2}\right) \right] ,
\label{Pi-geral}
\end{eqnarray}
with $h(D,2)=\left( 2\pi \right) ^{4-D/2}$, which, replaced in Eq.~(\ref
{lambda}), gives the effective boundary-dependent coupling constant in the
large-$N$ limit.

\subsection{Infrared behavior}

We can write Eq. (\ref{Pi-geral}) in the form 
\begin{equation}
\Pi (p,D,\{L_i\},m=0)= A(D)|p|^{D-4}+B_d(D,\{L_i\}),
\label{forma geral}
\end{equation}
with the coefficient of the $|p|$-term being 
\begin{equation}
A(D)=\left( 2\pi \right) ^{4-3D/2}2^{-D/2}b(D)\Gamma \left( 2-\frac D2%
\right) ,  \label{A(D)1}
\end{equation}
where we have defined 
\begin{equation}
b(D) =\int_0^1dx\,[x(1-x)]^{D/2-2}  =2^{3-D}\sqrt{\pi }\frac{\Gamma \left( \frac D2-1\right) }{\Gamma \left( 
\frac{D-1}2\right) },\qquad \text{for Re}(D)>2\text{,}  \label{def-b}
\end{equation}
and 
\begin{eqnarray}
B_d(D,\left\{ L_i\right\} ) &=&\frac{h(D,2)}{\left( 2\pi \right) ^4}%
\int_0^1dx\,\left[ \sum_{i=1}^d\sum_{n_i=1}^\infty \left( \frac{\sqrt{%
p^2x(1-x)}}{2\pi L_in_i}\right) ^{D/2-2}K_{D/2-2}\left( \frac 1{2\pi }\sqrt{%
p^2x(1-x)}L_in_i\right) \right.  \nonumber \\
&&\left. +2\sum_{i<j=1}^d\sum_{n_i,n_j=1}^\infty \left( \frac{\sqrt{p^2x(1-x)%
}}{2\pi \sqrt{L_i^2n_i^2+L_j^2n_j^2}}\right) ^{D/2-2}K_{D/2-2}\left( \frac 1{%
2\pi }\sqrt{p^2x(1-x)}\sqrt{L_i^2n_i^2+L_j^2n_j^2}\right) +\cdots \right. 
\nonumber \\
&&\left. +2^{d-1}\sum_{n_1,\ldots ,n_d=1}^\infty \left( \frac{\sqrt{p^2x(1-x)%
}}{2\pi \sqrt{L_1^2n_1^2+\cdots +L_d^2n_d^2}}\right) ^{D/2-2}K_{D/2-2}\left( 
\frac 1{2\pi }\sqrt{p^2x(1-x)}\sqrt{L_1^2n_1^2+\cdots +L_d^2n_d^2}\right)
\right] .  \nonumber  \label{Bd} \\
&&  \label{Bd}
\end{eqnarray}
We remark that, for the physically interesting dimension $D=3$, $b(3)=\pi $.
This implies that $A(3)=\pi /4$.

If an infrared-stable fixed point exists for any of the models with $d$
confining dimensions, it would be possible to determine it by a study of the
infrared behavior of the Callan--Symanzik $\beta $ function, {\em i.e.}, in
the neighborhood of $|p|=0$. Therefore, we should investigate the above
equations for $|p|\approx 0$.

In this case, we consider a typical term in Eq.~(\ref{Bd}), which has the
form 
\begin{equation}
\sum_{n_1,\ldots ,n_p=1}^\infty \left( \frac{\sqrt{p^2x(1-x)}}{2\pi \sqrt{%
L_1^2n_1^2+\cdots +L_p^2n_p^2}}\right) ^{D/2-s}K_{D/2-s}\left( \frac 1{2\pi }%
\sqrt{p^2x(1-x)}\sqrt{L_1^2n_1^2+\cdots +L_p^2n_p^2}\right) ,
\label{typical}
\end{equation}
with $s=2$ and $p=1,2,\ldots ,d$. In the $|p|\approx 0$ limit, we may use an
asymptotic formula for small values of the argument of the modified Bessel
functions \cite{abramowitz}, 
\begin{equation}
K_\nu (z)\approx \frac 12\Gamma (\nu )\left( \frac z2\right) ^{-\nu
}\;\;\;(z\sim 0,\;\;\;\text{Re}(\nu )>0)  \label{K}
\end{equation}
and Eq. (\ref{typical}) reduces to 
\begin{equation}
\frac 12\Gamma \left( \frac D2-s\right) E_p\left( \frac D2-s;L_1,\ldots
,L_p\right) .  \label{G2}
\end{equation}
It is expressed in terms of one of the multidimensional Epstein zeta
functions $E_p\left( \frac D2-s;L_1,\ldots ,L_p\right) $, for $p=1,2,\ldots
,d$, which are defined by~\cite{kirsten} 
\begin{equation}
E_p\left( \nu ;\sigma _1,\ldots ,\sigma _p\right) =\sum_{n_1,\ldots
,n_p=1}^\infty \left[ \sigma _1^2n_1^2+\cdots +\sigma _p^2n_p^2\right]
^{\,-\nu }\;.  \label{EfuncS}
\end{equation}
Notice that, for $p=1$, $E_p$ reduces to the Riemann zeta function $\zeta
(z)=\sum_{n=1}^\infty n^{-z}$. We then see from (\ref{G2}) that in this
limit the $p^2$-dependence of the modified Bessel functions exactly
compensates the one coming from the accompanying factors. Thus the remaining 
$p^2$-dependence is only that of the first term of (\ref{Pi-geral}), which
is the same for all number of compactified dimensions $d$.

One can also construct analytical continuations and recurrence relations for
the multidimensional Epstein functions, which permit to write them in terms
of modified Bessel and Riemann zeta functions \cite{ademir,kirsten}. One
gets 
\begin{eqnarray}
E_p\left( \nu ;L_1,\ldots ,L_p\right) &=&-\,\frac 1{2\,p}\sum_{i=1}^pE_{p-1}%
\left( \nu ;\ldots ,\widehat{L_i},\ldots \right) +\,\frac{\sqrt{\pi }}{%
2\,d\,\Gamma (\nu )}\Gamma \left( \nu -\frac 12\right) \sum_{i=1}^p\frac 1{%
L_i}E_{p-1}\left( \nu -\frac 12;\ldots ,\widehat{L_i},\ldots \right) 
\nonumber \\
&&+\frac{2\sqrt{\pi }}{p\,\Gamma (\nu )}W_p\left( \nu -\frac 12,L_1,\ldots
,L_p\right) \;,  \label{Wd}
\end{eqnarray}
where the hat over the parameter $L_i$ in the functions $E_{p-1}$ means that
it is excluded from the set $\{L_1,\ldots ,L_p\}$ (the others being the $p-1$
parameters of $E_{p-1}$), and 
\begin{equation}
W_p\left( \nu ;L_1,\ldots ,L_p\right) =\sum_{i=1}^p\frac 1{L_i}%
\sum_{n_1,...,n_p=1}^\infty \left( \frac{\pi n_i}{L_i\sqrt{\cdots +\widehat{%
L_in_i^2}+\cdots }}\right) ^\nu K_\nu \left( \frac{2\pi n_i}{L_i}\sqrt{%
\cdots +\widehat{L_in_i^2}+\cdots }\right) \;,  \label{WWd}
\end{equation}
with $\cdots +\widehat{L_in_i^2}+\cdots $ representing the sum $%
\sum_{j=1}^pL_j^2n_j^2\,-\,L_i^2n_i^2$.

We can derive expressions for each particular value of $d$, from 1 to $D$,
but let us restrict ourselves to the most expressive values, $d=1,2,3$. For $
D=3$, these correspond respectively to materials in the form of a film, a
wire, or a grain.

\subsubsection{One compactified dimension (a film)}

By taking $d=1$, the compactification of just one dimension, let us say,
along the $x_1$-axis, we are considering that the system is confined between
two planes, separated by a distance $L_1=L$. Physically, for $D=3$, this
corresponds to a film of thickness $L$. Then we have, from Eqs. (\ref{Bd}), (\ref{K})
and (\ref{EfuncS}), in the $|p|\approx 0$ limit, 
\begin{eqnarray}
B_{d=1}(D,L) &=&(2\pi )^{-D/2}\int_0^1dx\sum_{n=1}^\infty \left( \frac{\sqrt{%
p^2x(1-x)}}{2\pi nL}\right) ^{D/2-2}K_{D/2-2}\left( \frac 1{2\pi }nL\sqrt{%
p^2x(1-x)}\right)  \nonumber \\
&\sim &(2\pi )^{-D/2}2^{D/2-3}L^{4-D}\Gamma \left( \frac D2-2\right) \zeta (D-4),
\label{B-filme}
\end{eqnarray}
where $\zeta (z)$ is the Riemann zeta function. The above expression is
valid for all {\em odd} dimensions $D>5$, due to the poles of the $\Gamma $
and $\zeta $ functions. We can obtain an expression for smaller values of $D$
by using the recurrence relations, Eq.~(\ref{Wd}); in the present case, this
is equivalent to perform an analytic continuation of the Riemann zeta
function $\zeta (D-4)$ by means of its reflexion property \cite{abramowitz}, 
\begin{equation}
\zeta (z)=\frac{\Gamma \left( \frac{1-z}2\right) }{\Gamma (z/2)}\pi
^{z-1/2}\zeta (1-z),  \label{extensao}
\end{equation}
which gives 
\begin{equation}
\zeta (D-4)=\frac{\Gamma \left( \frac{5-D}2\right) }{\Gamma \left( \frac D2%
-2\right) }\pi ^{D-9/2}\zeta (5-D).  \label{extensao1}
\end{equation}
Then Eq. (\ref{B-filme}) becomes an expression valid for $2<D<4$ given by 
\begin{equation}
B_{d=1}(D,L)=2^{-3}\pi ^{(D-9)/2}L^{4-D}\Gamma \left( \frac{5-D}2\right)
\zeta (5-D).
\end{equation}
For $D=3$, we have $B_{d=1}(3,L)=L/48\pi $.

\subsubsection{Two compactified dimensions (a wire)}

Let us now take the case $d=2$, in which the system is confined
simultaneously between two parallel planes a distance $L_1$ apart from one
another normal to the $x_1$-axis and two other parallel planes, normal to
the $x_2$-axis separated by a distance $L_2$. That is, in the physical space
the material is bounded within an infinite wire of rectangular cross section 
$L_1\times L_2$. We then get for $|p|\approx 0$, 
\begin{eqnarray}
B_{d=2}(D;L_1,L_2) &=&(2\pi )^{-D/2}\left[ \int_0^1dx\sum_{n=1}^\infty
\left( \frac{\sqrt{p^2x(1-x)}}{2\pi nL_1}\right) ^{D/2-2}K_{D/2-2}\left( 
\frac 1{2\pi }nL_1\sqrt{p^2x(1-x)}\right) \right.  \nonumber \\
&&+\int_0^1dx\sum_{n=1}^\infty \left( \frac{\sqrt{p^2x(1-x)}}{2\pi nL_2}%
\right) ^{D/2-2}K_{D/2-2}\left( \frac 1{2\pi }nL_2\sqrt{p^2x(1-x)}\right) 
\nonumber \\
&&\left. +2\int_0^1dx\sum_{n_1,n_2=1}^\infty \left( \frac{\sqrt{p^2x(1-x)}}{%
2\pi \sqrt{L_1^2n_1^2+L_2^2n_2^2}}\right) ^{D/2-2}K_{D/2-2}\left( \frac 1{%
2\pi }\sqrt{L_1^2n_1^2+L_2^2n_2^2}\sqrt{p^2x(1-x)}\right) \right]  \nonumber
\label{Pi-fio} \\
&\sim &2^{-3}\pi ^{(D-9)/2}\left( L_1^{4-D}+L_2^{4-D}\right) \Gamma \left( \frac{%
5-D}2\right) \zeta (5-D)+2^{-2}\pi ^{-D/2}\Gamma \left( \frac D2-2\right)
E_2\left( \frac D2-2;L_1,L_2\right) ,  \nonumber  \label{Pi-fio} \\
&&  \label{Pi-fio}
\end{eqnarray}
with $E_2$ defined in Eq. (\ref{EfuncS}) and valid for Re$\left( D\right) >3.$

In particular, noticing that $E_1\left( \nu ;L_j\right) =L_j^{-2\nu }\zeta
(2\nu )$, one finds 
\begin{eqnarray}
E_2\left( \frac{D-2}2;L_1,L_2\right) &=&-\frac 14\left( \frac 1{L_1^{D-2}}+%
\frac 1{L_2^{D-2}}\right) \zeta (D-2)  \nonumber \\
&&+\;\frac{\sqrt{\pi }\Gamma (\frac{D-3}2)}{4\Gamma (\frac{D-2}2)}\left( 
\frac 1{L_1L_2^{D-3}}+\frac 1{L_1^{D-3}L_2}\right) \zeta (D-3)+\frac{\sqrt{%
\pi }}{\Gamma (\frac{D-2}2)}W_2\left( \frac{D-3}2;L_1,L_2\right) \;,
\label{Z1}
\end{eqnarray}
which is a meromorphic function of $D$, symmetric in the parameters $L_1$
and $L_2$. The function $W_2\left( (D-3)/2;L_1,L_2\right) $ in Eq.~(\ref{Z1}%
) is the particular case of Eq.~(\ref{WWd}) for $p=2$.

This equation presents no problems for $3<D<4$ but, for $D=3$, the first and
second terms between brackets of Eq. (\ref{Z1}) are divergent due to the $%
\zeta $ function and the $\Gamma $ function, respectively. However, these
two divergences cancel out. No regularization is
needed. This can be seen by remembering the property 
\begin{equation}
\lim_{z\rightarrow 1}\left[ \zeta (z)-\frac 1{z-1}\right] =\gamma \;,
\label{extensao1}
\end{equation}
where $\gamma \approx 0.5772$ is the Euler--Mascheroni constant and, using
the expansion of $\Gamma ((D-3)/2)$ around $D=3$, 
\begin{equation}
\Gamma \left( \frac{D-3}2\right) \approx \frac 2{D-3}+\Gamma ^{\prime }(1)\,,
\label{GD3}
\end{equation}
$\Gamma ^{\prime }(z)$ standing for the derivative of the $\Gamma $ function
with respect to $z$. For $z=1$, it coincides with the Euler digamma function 
$\psi (1)$, which has the particular value $\psi (1)=-\gamma $. The two
divergent terms generated by the use of formulas (\ref{extensao1}) and (\ref
{GD3}) cancel exactly for $D=3$. Thus, remembering Eq.~(\ref{def-b}), the
domain of existence of $B_{d=2}(D;L_1,L_2)$ can, as in the case of films, be
extended to $2<D<4$.

\subsubsection{Three compactified dimensions (a grain)}

Finally, we may compactify three of the dimensions, which leaves us in $D=3$
with a system which is a grain of some material in the form of a
parallelepiped. We have, for arbitrary $D$, for $|p|\approx 0$, 
\begin{eqnarray}
B_{d=3}(D;L_1,L_2,L_3) &=&(2\pi )^{-D/2}\left[ \int_0^1dx\sum_{n=1}^\infty
\left( \frac{\sqrt{p^2x(1-x)}}{2\pi nL_1}\right) ^{D/2-2}K_{D/2-2}\left( 
\frac 1{2\pi }nL_1\sqrt{p^2x(1-x)}\right) \right.  \nonumber \\
&&+\int_0^1dx\sum_{n=1}^\infty \left( \frac{\sqrt{p^2x(1-x)}}{2\pi nL_2}%
\right) ^{D/2-2}K_{D/2-2}\left( \frac 1{2\pi }nL_2\sqrt{p^2x(1-x)}\right) 
\nonumber \\
&&+\int_0^1dx\sum_{n=1}^\infty \left( \frac{\sqrt{p^2x(1-x)}}{2\pi nL_3}%
\right) ^{D/2-2}K_{D/2-2}\left( \frac 1{2\pi }nL_3\sqrt{p^2x(1-x)}\right) 
\nonumber \\
&&+2\int_0^1dx\sum_{n_1,n_2=1}^\infty \left( \frac{\sqrt{p^2x(1-x)}}{2\pi 
\sqrt{L_1^2n_1^2+L_2^2n_2^2}}\right) ^{D/2-2}K_{D/2-2}\left( \frac 1{2\pi }%
\sqrt{L_1^2n_1^2+L_2^2n_2^2}\sqrt{p^2x(1-x)}\right)  \nonumber \\
&&+2\int_0^1dx\sum_{n_1,n_3=1}^\infty \left( \frac{\sqrt{p^2x(1-x)}}{2\pi 
\sqrt{L_1^2n_1^2+L_3^2n_3^2}}\right) ^{D/2-2}K_{D/2-2}\left( \frac 1{2\pi }%
\sqrt{L_1^2n_1^2+L_3^2n_3^2}\sqrt{p^2x(1-x)}\right)  \nonumber \\
&&+2\int_0^1dx\sum_{n_2,n_3=1}^\infty \left( \frac{\sqrt{p^2x(1-x)}}{2\pi 
\sqrt{L_2^2n_2^2+L_3^2n_3^2}}\right) ^{D/2-2}K_{D/2-2}\left( \frac 1{2\pi }%
\sqrt{L_2^2n_2^2+L_3^2n_3^2}\sqrt{p^2x(1-x)}\right)  \nonumber \\
&&+4\int_0^1dx\sum_{n_1,n_2,n_3=1}^\infty \left( \frac{\sqrt{p^2x(1-x)}}{%
2\pi \sqrt{L_1^2n_1^2+L_2^2n_2^2+L_3^2n_3^2}}\right) ^{D/2-2}  \nonumber \\
&&\left. \qquad \qquad \qquad \qquad \times K_{D/2-2}\left( \frac 1{2\pi }%
\sqrt{L_1^2n_1^2+L_2^2n_2^2+L_3^2n_3^2}\sqrt{p^2x(1-x)}\right) \right] 
\nonumber  \label{Pi-grao} \\
&\sim &\frac 18\pi ^{(D-9)/2}\left( L_1^{4-D}+L_2^{4-D}+L_3^{4-D}\right) \Gamma
\left( \frac{5-D}2\right) \zeta (5-D)  \nonumber \\
&&+\frac 1{4\pi ^{D/2}}\Gamma \left( \frac D2-2\right) \left[ E_2\left( 
\frac D2-2;L_1,L_2\right) +E_2\left( \frac D2-2;L_1,L_3\right) +E_2\left( 
\frac D2-2;L_2,L_3\right) \right]  \nonumber \\
&&\left. +\frac 1{2\pi ^{D/2}}\Gamma \left( \frac D2-2\right) E_3\left( 
\frac D2-2;L_1,L_2,L_3\right) \right] .
\end{eqnarray}

The analytical structure of the function $E_3\left(
(D-2)/2;L_1,L_2,L_3\right) $ in the equation above can be obtained from the
general symmetrized recurrence relation given by Eqs.~(\ref{Wd}) and (\ref
{WWd}); explicitly, one has 
\begin{eqnarray}
E_3\left( \frac{D-2}2;L_1,L_2,L_3\right) &=&-\frac 16\sum_{i<j=1}^3E_2\left( 
\frac{D-2}2;L_i,L_j\right) +\frac{\sqrt{\pi }\Gamma \left( \frac{D-3}2%
\right) }{6\Gamma \left( \frac{D-2}2\right) }\sum_{i,j,k=1}^3\frac{%
(1+\varepsilon _{ijk})}2\frac 1{L_i}E_2\left( \frac{D-2}2;L_j,L_k\right) 
\nonumber \\
&&+\frac{2\sqrt{\pi }}{3\Gamma \left( \frac{D-2}2\right) }\,W_3\left( \frac{%
D-3}2;L_1,L_2,L_3\right) ,  \label{E3}
\end{eqnarray}
where $\varepsilon _{ijk}$ is the totally antisymmetric symbol and the
function $W_3$ is a particular case of Eq.~(\ref{WWd}). The first two terms
in the square bracket of Eq.~(\ref{E3}) diverge as $D\rightarrow 3$ due to
the poles of the $\Gamma $ and $\zeta $ functions. However, as it happens in
the case of wires, it can be shown that these divergences cancel exactly one
another, leaving an extended domain of validity $2<D<4$, for $%
B_{d=3}(D;L_1,L_2,L_3)$.

\subsection{The $\beta $ function and the fixed points}

For all $d\leq D$, within the domain of validity of $D$, we have, by
inserting (\ref{forma geral}) in Eq. (\ref{lambda}), the running coupling
constant 
\begin{equation}
\lambda \left( |p|\approx 0,D,\{L_i\}\right) \approx \frac \lambda {%
1+\lambda \left[ A(D)|p|^{D-4}+B_d\left( D,\left\{ L_i\right\} \right)
\right] }.  \label{g3}
\end{equation}
Let us take $|p|$ as a running scale, and define the dimensionless coupling 
\begin{equation}
g=\lambda \left( p,D,\{L_i\}\right) |p|^{D-4}.  \label{g1}
\end{equation}
We recall that in the previous expressions $p$ is a $D$-dimensional vector.

It is widely known that the $\beta $ function controls the rate of the
renormalization-group flow of the running coupling constant and that a
(nontrivial) fixed point of this flow is given by a (nontrivial) zero of the 
$\beta $ function. For $|p|\approx 0$, it is obtained straightforwardly from
Eq. (\ref{g1}): 
\begin{equation}
\beta (g)=|p|\frac{\partial g}{\partial |p|}\approx (D-4)\left[
g-A(D)g^2\right] ,  \label{beta}
\end{equation}
from which we get the infrared-stable fixed point 
\begin{equation}
g_{*}(D)=\frac 1{A(D)}.  \label{gstar}
\end{equation}
We see that the $L_i$-dependent $B_d$-part of the subdiagram $\Pi $ does not
play any role in this expression and, as remarked before, $A(D)$ is the same
for all number of compactified dimensions, so is $g_{*}$ only dependent on
the space dimension.

\section{The system with an external magnetic field}

\subsection{The Landau-level basis}

In this section, we take the same $N$-component Ginzburg--Landau model to
describe the behavior of confined systems, now in the presence of an
external magnetic field, at leading order in $1/N$. The system is again
constrained to a $d$-dimensional subspace of ${\rm{\bf  R} }^D$ in the form of a
parallelepiped. The Hamiltonian density is then modified to 
\begin{equation}
{\cal H}=\left[ \left( \partial _\mu -ieA_\mu ^{\text{ext}}\right) \varphi
_a\right] \left[ \left( \partial ^\mu -ieA^{\text{ext,}\mu }\right) \varphi
_a\right] +m^2\varphi _a\varphi _a+u\,(\varphi _a\varphi _a)^2,
\label{hamiltoniana2}
\end{equation}
where summation over repeated indices is assumed and $m^2=\alpha (T-T_c)$,
with $\alpha >0$. For $D=3$, from a physical point of view, such Hamiltonian
is supposed to describe type-II superconductors. In this case, we assume
that the external magnetic field ${\bf H}$ is parallel to the $z$-axis and
we choose the gauge ${\bf A}^{\text{ext}}=(0,xH,0)$. The model with $N $
complex components is taken in the large-$N$ limit with $Nu=\lambda $ fixed.
If we consider the system in unlimited space, the field $\varphi $ should be
written in terms of the well-known Landau-level basis, 
\begin{equation}
\varphi ({\bf r})=\sum_{\ell =0}^\infty \int \frac{dp_y}{2\pi }\int \frac{%
d^{D-2}p}{\left( 2\pi \right) ^{D-2}}\tilde{\varphi}_{\ell ,p_y,{\bf p}}\chi
_{\ell ,p_y,{\bf p}}({\bf r}),
\end{equation}
where $\chi _{\ell ,p_y,{\bf p}}({\bf r})$ are the Landau-level
eigenfunctions given by 
\begin{equation}
\chi _{\ell ,p_y,{\bf p}}({\bf r})=\frac 1{\sqrt{2^\ell }\ell !}\left( \frac 
\omega \pi \right) ^{1/4}e^{i\left( {\bf p}\cdot {\bf r}+p_yy\right)
}e^{-\omega (x-p_y/\omega )^2/2}H_\ell \left( \sqrt{\omega }x-\frac{p_y}{%
\sqrt{\omega }}\right) ,
\end{equation}
with energy eigenvalues $E_\ell \left( \left| {\bf p}\right| \right) =$ $%
\left| {\bf p}\right| ^2+\left( 2\ell +1\right) \omega +m^2$ and $\omega =eH$
is the so-called cyclotron frequency. In the above equation, ${\bf p} $ and $%
{\bf r}$ are ($D-2$)-dimensional vectors.

Let us consider the system confined as in the previous sections, and use
Cartesian coordinates ${\bf r}=(x_1,\ldots ,x_d,{\bf z})$, where ${\bf z}$
now is a ($D-2-d$)-dimensional vector, with corresponding momenta ${\bf k}%
=(k_1,\ldots ,k_d,{\bf q})$, ${\bf q}$ being a $(D-2-d)$-dimensional vector
in momentum space. That is, the superconducting material is confined to a
subspace of the $D$-dimensional Euclidean space in the form of a $d$%
-dimensional parallelepiped. Under these conditions, the generating
functional of correlation functions is written as 
\begin{equation}
{\cal Z}=\int {\cal D}\varphi ^{*}{\cal D}\varphi \,\exp \left(
-\int_0^{L_1}dx_1\cdots \int_0^{L_d}dx_d\int d^{D-d-2}z\;{\cal H}(|\varphi
|,|\nabla \varphi |\right) ,
\end{equation}
with the field $\varphi (x_1,\ldots ,x_d,{\bf z})$ satisfying the
box-confinement condition as in Section II. Then the field representation
should be modified and have a mixed series-integral Fourier expansion of the
form 
\begin{equation}
\varphi (x_1,\ldots ,x_d,{\bf z})=\sum_{\ell =0}^\infty
\sum_{i=1}^d\sum_{n_i=-\infty }^\infty c_{n_i}\int \frac{dp_y}{2\pi }\int
d^{D-d-2}{\bf q}\;b({\bf q})e^{-i\omega _{n_i}x\;-i{\bf q}\cdot {\bf z}}%
\tilde{\varphi}_\ell (\omega _{n_i},{\bf q}),  \label{Fourier2}
\end{equation}
where, for $i=1,\ldots ,d$, $\omega _{n_i}=2\pi n_i/L_i$ and the
coefficients $c_{n_i}$ and $b({\bf q})$ correspond respectively to the
Fourier series representation over the $x_i$ and to the Fourier integral
representation over the ($D-d-2)$-dimensional ${\bf z}$-space. As was done
previously, we now apply the Matsubara-like formalism according to (\ref
{Matsubara}).

\subsection{Infrared behavior}

In the following, we consider only the lowest Landau level $\ell =0$. For $%
D=3$, this assumption usually corresponds to the description of
superconductors in the extreme type-II limit. Under this assumption, we
obtain that the effective $\left| \varphi \right| ^4$ interaction in
momentum space and at the critical point as 
\begin{equation}
\lambda (p,D,\{L_i\};\omega )=\frac \lambda {1+\lambda \omega e^{-(1/2\omega
)(p_1^2+p_2^2)}\Pi (p,D,\left\{ L_i\right\} ,\,m=0;\omega )},  \label{novoU}
\end{equation}
where the single 1-loop bubble $\Pi (p,D,\{L_i\},m=0;\omega )$ is given by 
\begin{eqnarray}
\Pi (p,D,\{L_i\},m=0;\omega ) &=&\frac 1{L_1\cdots L_d}\sum_{i=1}^d%
\sum_{n_i=-\infty }^\infty \int_0^1dx\int \frac{d^{D-d-2}q}{(2\pi )^{D-d-2}}
\nonumber \\
&&\times \frac 1{\left[ {\bf q}^2+\omega _{n_1}^2+\cdots +\omega
_{n_d}^2+p^2x(1-x)\right] ^2}.  \nonumber \\
&&  \label{novoPi}
\end{eqnarray}
This is the same kind of expression that is encountered in the previous
section, Eq. (\ref{Pi2}), with the only modification that $D\rightarrow D-2$%
. Also, one should be reminded that $p$ is now a ($D-2$)-dimensional vector.
The analysis is then performed along the same lines and we obtain,
analogously, 
\begin{eqnarray}
\Pi (p,D,\{L_i\}, &&m=0;\omega )=\left( 2\pi \right) ^{1-D/2}\,\left[
2^{1-D/2}\frac 1{\left( 2\pi \right) ^2}c(D)\Gamma \left( 3-\frac D2\right)
\left( p^2\right) ^{D/2-3}\right.  \nonumber \\
&&+\int_0^1dx\sum_{i=1}^d\sum_{n_i=1}^\infty \left( \frac{\sqrt{p^2x(1-x)}}{%
2\pi L_in_i}\right) ^{D/2-3}K_{D/2-3}\left( \frac 1{2\pi }\sqrt{p^2x(1-x)}%
L_in_i\right)  \nonumber \\
&&+2\int_0^1dx\sum_{i<j=1}^d\sum_{n_i,n_j=1}^\infty \left( \frac{\sqrt{%
p^2x(1-x)}}{2\pi \sqrt{L_i^2n_i^2+L_j^2n_j^2}}\right)
^{D/2-3}K_{D/2-3}\left( \frac 1{2\pi }\sqrt{p^2x(1-x)}\sqrt{%
L_i^2n_i^2+L_j^2n_j^2}\right) +\cdots  \nonumber \\
&&+2^{d-1}\int_0^1dx\sum_{n_1,\ldots ,n_d=1}^\infty \left( \frac{\sqrt{%
p^2x(1-x)}}{2\pi \sqrt{L_1^2n_1^2+\cdots +L_d^2n_d^2}}\right) ^{D/2-3} 
\nonumber \\
&&\left. \qquad \qquad \qquad \qquad \times K_{D/2-3}\left( \frac 1{2\pi }%
\sqrt{p^2x(1-x)}\sqrt{L_1^2n_1^2+\cdots +L_d^2n_d^2}\right) \right] , 
\nonumber \\
&&  \label{novoPi-geral}
\end{eqnarray}
where 
\begin{equation}
c(D) =\int_0^1dx\left( x(1-x)\right) ^{D/2-3}  \label{def-b2} 
=2^{5-D}\sqrt{\pi }\frac{\Gamma \left( \frac D2-2\right) }{\Gamma \left( 
\frac{D-3}2\right) },\qquad \text{for Re}(D)>4\text{.}  \label{def-b2}
\end{equation}

As for the infrared behavior of the $\beta $ function, it suffices to study
it in the neighborhood of $\left| p\right| =0$, so that we can again use the
asymptotic formula (\ref{K}). It turns out that in the $\left| p\right|
\approx 0$ limit, the bubble $\Pi $ is written in the form 
\begin{equation}
\Pi (|p|\approx 0,D,\{L_i\},m=0;\omega )=A_1(D)\left| p\right|
^{D-6}+C_d(D,\left\{ L_i\right\} ),
\end{equation}
with 
\begin{equation}
A_1(D)=\left( 2\pi \right) ^{-D/2-1}2^{1-D/2}c(D)\Gamma \left( 3-\frac D2%
\right) ,  \label{A(D)2}
\end{equation}
and where the quantity $C_d(D,\left\{ L_i\right\} )$ is obtained by simply
making the change $D\rightarrow D-2$ in the formula for $B_d(D,\left\{
L_i\right\} )$ in the preceding section.

\subsection{Fixed points}

Let us define a dimensionless coupling constant by 
\[
g=\omega \lambda (|p|\approx 0,D,\{L_i\};\omega )\left| p\right| ^{D-6}. 
\]
Then, after performing manipulations entirely analogous to those in Section
and recalling Eq.~(\ref{def-b2}), we have the extended domain of validity $%
4<D<6$ for the quantities $C_{d=1}(D;L_1)$, $C_{d=2}(D;L_1,L_2)$ and $%
C_{d=3}(D;L_1,L_2,L_3)$.

As in the preceding section, we take as a running scale $|p|$, and define
the dimensionless coupling 
\begin{equation}
g^{(1)}=\omega \lambda (p_1=p_2=0,D,\{L_i\})|p|^{D-6},  \label{g1(1)}
\end{equation}
where we remember that in this context $p$ is a $(D-2)$-dimensional vector.
Then, we obtain the $\beta $ function for $|p|\approx 0$: 
\begin{equation}
\beta (g)=|p|\frac{\partial g^{(1)}}{\partial |p|}\approx (D-6)\left[
g^{(1)}-A_1(D)g^{(1)2}\right] ,  \label{beta2}
\end{equation}
from which the infrared-stable fixed point 
\begin{equation}
g_{*}^{(1)}(D)=\frac 1{A_1(D)}  \label{gstar1}
\end{equation}
is obtained.

\section{Concluding remarks}

In this article, we have discussed the infrared behavior and the fixed-point structure of the $N$%
-component Ginzburg--Landau model in the large-$N$ limit, the system being
confined in a $d$-dimensional box with edges of length $L_i$, $i=1,2,\ldots
,d$ (compactification in a $d$-dimensional subspace). For $D=3$ and $d=1,2,3$%
, the system is supposed to describe, respectively, a film of thickness $L$,
an infinitely long wire of cross-section $L_1\times L_2$, and a grain of
volume $L_1\times L_2\times L_3$. We have studied the cases in which the
system has no external influence and in which the system is submitted to the
action of an applied external magnetic field. In both situations, with or
without an external magnetic field, we get the result that the existence of
an infrared-stable fixed point depends only on the space dimension $D$; it
does not depend on the number of compactified dimensions.

In the absence of an external magnetic field, we find that, for $2<D<4$, our
result is the existence of an infrared-stable fixed point, in agreement with
previous renormalization-group calculations for materials in bulk form (all $%
L_i=\infty $) in the literature (see, for instance, \cite{zinn} and other
references therein). Taking $D=3$, we demonstrate directly that in the
absence of a magnetic field, the superconducting transition in films, wires
and grains is a second-order one. Moreover, the fixed point is independent
of the size of the system or, in other words, the nature of the transition
in the absence of a magnetic field is insensitive to the confining geometry.

In the case of the system in the presence of an external magnetic field, it
is interesting to compare our results with those obtained for type-II
materials in bulk form. For instance, a large-$N$ analysis and a functional
renormalization-group study performed in Refs. \cite{moore,radz,moore1}
conclude for a second-order transition in dimensions $4<D<6$. The same
conclusion is obtained in Ref. \cite{flavio}. The authors of Ref. \cite
{moore} claim, moreover, that the inclusion of fluctuations does not alter
significantly the main characteristic of the system, that is, the existence
of a continuous transition into a spatially homogeneous condensate. For the
system under the action of an external magnetic field, the existence of a
fixed point for $4<D<6$ should be taken as an indication, not as a
demonstration, of the existence of a continuous transition. As already
discussed in~\cite{moore,moore1}, in this case, even if infrared fixed
points exist, none of them can be completely attractive. The existence of an
infrared fixed point in the presence of a magnetic field, as found in this
paper, does not assure the (formal) existence of a second-order transition.
Anyway, we conclude that, for materials in the form of films, wires and
grains under the action of an external magnetic field, as is also the case
for materials in bulk form, if there exists a phase transition for $D<4$, in
particular in $D=3$, it should not be a second-order one.

Our results for a confined material in absence of an external magnetic field
in dimension $2<D<4$ and for confined materials submitted to an external
magnetic field for dimensions $4<D<6$ are in agreement with previous results
for the bulk. Notice the shift of $2$ in the range of dimensions for which a
second-order transition would occur in both cases. As a final remark, from
Eqs.~(\ref{def-b}) and (\ref{def-b2}), we see that, for $D\leq 2$ and for $%
D\leq 4$, respectively, in particular for $D=1$, severe infrared divergences
appear under the form of a divergence of the integrals in the quantities $%
b(D)$ and $c(D)$. Here we recover the well-known Peierls theorem, which
forbids the existence of phase transitions in unidimensional spaces.

\section{Acknowledgments}

A.P.C.M. acknowleges support by CNPq and FAPERJ (Brazil) and hospitality from CERN - Th. Division during october 2010.


\begin{references}
\bibitem{affleck}  I. Affleck and E. Br\'{e}zin, Nucl. Phys. B {\bf 257},
451 (1985).

\bibitem{lawrie}  I. D. Lawrie, Phys. Rev. B {\bf 50}, 9456 (1994).

\bibitem{lawrie1}  I. D. Lawrie, Phys. Rev. Lett. {\bf 79}, 131 (1997).

\bibitem{brezin}  E. Br\'{e}zin, D. R. Nelson and A. Thiaville, Phys. Rev. B 
{\bf 31}, 7124 (1985).

\bibitem{radz}  L. Radzihovsky, Phys. Rev. Lett. {\bf 74}, 4722 (1995);{\bf 
\ 76}, 4451 (1996); I. F. Herbut and Z. Tesanovic, {\em ibid.} {\bf 76},
4450 (1996).

\bibitem{flavio}  C. de Calan, A. P. C. Malbouisson and F. S. Nogueira,
Phys. Rev. B {\bf 64}, 212502 (2001).

\bibitem{malbo}  A. P. C. Malbouisson, Phys. Rev. B {\bf 66}, 092502 (2002).

\bibitem{zinn}  J. Zinn-Justin, {\em Quantum Field Theory and Critical
Phenomena}, 3rd edition (Clarendon Press, Oxford, 1996).

\bibitem{cardy}  J. L. Cardy (ed.), {\em Finite Size Scaling} (North
Holland, Amsterdam, 1988).

\bibitem{livro}  F. C. Khanna, A. P. C. Malbouisson, J. M. C. Malbouisson
and A. E. Santana, {\em Thermal Quantum Field Theory - Algebraic Aspects and
Applications} (World Scientific, Singapore, 2009).

\bibitem{fosco}  C. D. Fosco and A. Lopez, Nucl. Phys. B {\bf 538}, 685
(1999).

\bibitem{fadolfo}  L. Da Rold, C. D. Fosco and A. P. C. Malbouisson, Nucl.
Phys. B {\bf 624}, 485 (2002).

\bibitem{polchinski}  J. Polchinski, Commun. Math. Phys. {\bf 104}, 37
(1986).

\bibitem{atickwitten}  J. J. Atick and E. Witten, Nucl. Phys. B {\bf 310},
291 (1988).

\bibitem{tashk}  A. E. Santana, J. M. C. Malbouisson, A. P. C. Malbouisson
and F. C. Khanna, ``Thermal Field Theory: Algebraic Aspects and Applications
to Confined Systems'', in {\em Non-Linear Dynamics and Fundamental
Interactions}, eds. F. Khanna and D. Matrasulov (Springer, New York, 2005).

\bibitem{ademir}  A. P. C. Malbouisson, J. M. C. Malbouisson and A. E.
Santana, Nucl. Phys. B {\bf 631}, 83 (2002).

\bibitem{khanna1}  F. C. Khanna, A. P. C. Malbouisson and A. E. Santana,
Ann. Phys. (N.Y.) {\bf 324}, 1931 (2009).

\bibitem{ford1}  N. D. Birrell and L. H. Ford, Phys. Rev. D {\bf 22}, 330
(1974).

\bibitem{hebe}  H. Queiroz, J. C. da Siva, F. C. Khanna, J. M. C.
Malbouisson, M. Revzen and A. E. Santana, Ann. Phys. (N.Y.) {\bf 317}, 220
(2005).

\bibitem{jura1}  J. C. da Silva, F.C. Khanna, A. Matos Neto and A. E.
Santana, Phys. Rev. A {\bf 66}, 052101 (2002).

\bibitem{mostep3}  M. Bordag, U. Mohideed and V. M. Mostepanenko, {\em New
Developments in Casimir Effect}, Phys. Rep. {\bf 353}, 1 (2001).

\bibitem{amms}  L. M. Abreu, A. P. C. Malbouisson, J. M. C. Malbouisson and
A. E. Santana, Phys. Rev. B {\bf 67}, 212502 (2003).

\bibitem{linhares}  C. A. Linhares, A. P. C. Malbouisson, Y. W. Milla and I.
Roditi, Phys. Rev. B {\bf 73}, 214525 (2006).

\bibitem{linhares1}  C. A. Linhares, A. P. C. Malbouisson, Y. W. Milla and
I. Roditi, Eur. Phys. J. B {\bf 60}, 353 (2007).

\bibitem{jmp}  L. M. Abreu, C. de Calan, A. P. C. Malbouisson, J. M. C.
Malbouisson and A. E. Santana, J. Math. Phys. {\bf 46}, 012304 (2005).

\bibitem{jmp1}  A. P. C. Malbouisson, J. M. C. Malbouisson and R. C.
Pereira, J. Math. Phys. {\bf 50}, 083304 (2009).

\bibitem{prd11}  L.M. Abreu, A.P.C. Malbouisson and J.M.C. Malbouisson,
Phys. Rev. D {\bf 83}, 025001 (2011).

\bibitem{epl10}  F.C. Khanna, A.P.C. Malbouisson, J.M.C. Malbouisson and
A.E. Santana, Europhys. Lett. {\bf 92}, 11001 (2010).

\bibitem{gnn}  A. P. C. Malbouisson, J. M. C. Malbouisson, A. E. Santana and
J. C. Silva, Phys. Lett. B {\bf 583}, 373 (2004).

\bibitem{gnn2}  F. C. Khanna, A. P. C. Malbouisson, J. M. C. Malbouisson, T.
Rocha Filho, J. C. Silva and A. E. Santana, Phys. Lett. B {\bf 624}, 316
(2005).


\bibitem{npb09}  L. M. Abreu, A. P. C. Malbouisson, J. M. C. Malbouisson and
A. E. Santana, Nucl. Phys. B {\bf 819}, 127 (2009).

\bibitem{jmario}  A. P. C. Malbouisson and J. M. C. Malbouisson, J. Phys. A:
Math. Gen. {\bf 35}, 2263 (2002).

\bibitem{gino}  G. N. J. A\~{n}a\~{n}os, A. P. C. Malbouisson and N. F.
Svaiter, Nucl. Phys. B {\bf 547}, 221 (1999).

\bibitem{elizalde}  A. Elizalde and E. Romeo, J. Math. Phys. {\bf 30}, 1133
(1989).

\bibitem{abramowitz}  M. Abramowitz and I. A. Stegun (eds.), {\em Handbook
of Mathematical Functions} (Dover, New York, 1965).

\bibitem{kirsten}  K. Kirsten, J. Math. Phys. {\bf 35}, 459 (1994).

\bibitem{moore}  M. A. Moore, T. J. Newman, A. J. Bray and S.-K. Chin, Phys.
Rev. B {\bf 58}, 936 (1998).

\bibitem{moore1}  T. J. Newman and M. A. Moore, Phys. Rev. B {\bf 54}, 6661
(1996).
\end{references}
\end{document}